\documentclass{emulateapj}
\usepackage{graphicx}

\slugcomment{Draft Version March 24, 2016}
\shorttitle{High-z Radio Loud AGN Environments and Clustering}
\shortauthors{Kotyla et al.}

\begin{document}
\title{The Environment of $z >1$ 3CR Radio Galaxies and QSOs: From Proto-Clusters to clusters of Galaxies?}

\author{J.~P. Kotyla\altaffilmark{1}, M. Chiaberge\altaffilmark{1,2},  
S. Baum\altaffilmark{4}, A. Capetti\altaffilmark{5}, B. Hilbert\altaffilmark{1}, F.~D. Macchetto\altaffilmark{1}, G.~K. Miley\altaffilmark{6}, C.~P. O'Dea\altaffilmark{4}, E.~S. Perlman\altaffilmark{7},W.~B. Sparks\altaffilmark{1}, \& G.~R. Tremblay\altaffilmark{3}}

\email{marcoc@stsci.edu}
\altaffiltext{1}{Space Telescope Science Institute, 3700 San Martin Drive,Baltimore, MD 21218}
\altaffiltext{2}{Center for Astrophysical Sciences, Johns Hopkins University,3400 N. Charles Street, Baltimore, MD 21218, USA}
\altaffiltext{3}{Yale University, Department of Astronomy, 260 Whitney Ave, New Haven, CT 06511}
\altaffiltext{4}{University of Manitoba, Dept of Physics and Astronomy, 66 Chancellors Cir., Winnipeg, MB R3T 2N2 Canada}
\altaffiltext{5}{INAF-Osservatorio Astronomico di Torino, Via Osservatorio 20, 10025 Pino Torinese (TO), Italy}
\altaffiltext{6}{Universiteit Leiden, Rapenburg 70, 2311 EZ Leiden, Netherlands}
\altaffiltext{7}{Florida Institute of Technology, 150 W University Blvd, Melbourne, FL 32901}

\begin{abstract}
We study the cluster environment for a sample of 21 radio loud AGN from the 3CR catalog at $z>1$, 12 radio galaxies and 9 quasars with HST images in the optical and IR. We use two different approaches to determine cluster candidates. We identify the early type galaxies (ETGs) in every field by modeling each of the sources within a 40$"$ radius of the targets with a S\`{e}rsic profile. Using a simple passive evolution model, we derive the expected location of the ETGs on the red sequence (RS) in the color-magnitude diagram for each of the fields of our sources. For seven targets, the model coincides with the position of the ETGs. A second approach involves a search for over densities. We compare the object densities of the sample as a whole and individually against control fields taken from the GOODS-S region of 3D-HST survey. With this method we determine the fields of 10 targets to be cluster candidates. Four cluster candidates are found by both methods. The two methods disagree in some cases, depending on the specific properties of each field. For the most distant radio galaxy in the 3CR catalog (3C257 at $z = 2.47$), we identify a population of bluer ETGs that lie on the expected location of the RS model for that redshift. This appears to be the general behavior of ETGs in our fields and it is possibly a signature of the evolution of such galaxies. Our results are consistent with half of the z $>$ 1 radio galaxies being located in dense, rapidly evolving environments. 
\end{abstract}


\section{Introduction}

Radio galaxies (RGs) and radio loud quasars (QSOs) are among the most energetic phenomena in the universe. 
The hosts of these objects at low redshifts ($z < 0.3$) are massive ($M\sim10^{11}M_{\odot}$) giant elliptical galaxies \citep{zirbel96,donzelli07}. 
Their powerful jets are believed to be produced by rapidly spinning supermassive black holes \citep[SMBHs,][]{blanford77,colbert95,ghisellini14}.
There is also growing evidence that their black hole (BH) masses are above $\sim10^{8} M_\odot$ \citep{laor00, dunlop03, best05,chiaberge11,calderone13,castignani13,mao15}.
Radio loud (RL) active galactic nuclei (AGN) are typically found in rich Mpc-scale environments. At low redshifts $z<0.3$ the fraction of RGs that reside in clusters is as high as $70\%$ \citep{zirbel97} and a large fraction of them are located in the cluster cD galaxy.
At higher redshifts ($z \sim 0.5$ and above) the fraction of RL AGN in clusters is $\sim 50\%$ \citep{prestage88,hillLilly91,best00,galametz12, wylezalek13}. 
However, due to the lack of study of statistically meaningful samples for $z >1$, the exact fraction of RGs residing in clusters at these redshifts is still not firmly known.
This is in fact a central question not only for our understanding of the physics of the RL AGN phenomenon, but also for high-z clusters searches. In fact, high-z radio galaxies are often used as beacons for clusters or proto-clusters at $z>2$ \citep[][for a review]{miley08}.

A Hubble Space Telescope (HST) Cycle 20 snapshot program \citep[GO13023, P.I. Chiaberge, M.,][]{hilbert16} was designed to study the environment of a sample of 3CR \citep{spinrad85} RGs and QSOs at $z>1$ in much greater detail. 
In particular, one of the central goals of the project was to determine the fraction of clusters in the high z 3CR sample.   
Some of the most commonly used methods for determining the presence of clusters are based on the X-ray emission from the intracluster medium \citep{rosati02}, the Sunyaev Zel'dovich effect \citep{sunyaev72}, the red sequence (RS) method \citep{gladders_yee00}, as well as the search for over densities of galaxies through a number of statistical tools \citep[][and references therein]{castignani14}. The RS technique is known to identify clusters out to a redshift of $z \sim 2$. Such a method is based on a pattern found in color magnitude diagrams (CMD) due to the passive evolution of early-type galaxies.
  
In this paper we use two different approaches to study the Mpc scale environment of 3CR sources and determine their possible association with clusters or groups.  
Firstly, we focus on investigating the presence of a RS in the field of each target. 
Secondly, we compare the density of objects in each field against the average density of a control sample.  
The plan of the paper is as follows: in Section~\ref{data} we describe the sample and the HST observations; in Section~\ref{photometry} we discuss our method to detect objects, and perform photometry; Section~\ref{methods} focuses on detailing the methods we use to assess the presence of a cluster and describes the results; in Section~\ref{discussion} we discuss our findings. Lastly,  in Section~\ref{conclusion} we draw conclusions. 

Throughout the paper we use the AB magnitude system and a $\Lambda$CDM cosmological model with the following 
parameters: $H_{0} = 70$ km $s^{-1}$ $Mpc^{-1}$; $\Omega_{m} = 0.27$; $\Omega_{\Lambda} = 0.73$. 
\section{Observations and Data Reduction}
\label{data}
Our targets were observed in optical and near-IR with HST's
Wide Field Camera 3 (WFC3) between December 2012 and May 2013 as part of snapshot program GO13023. 
HST snapshot surveys of complete samples are well suitable for statistical studies, since the observations are scheduled by randomly picking objects from the original target list to fill gaps in the HST schedule.
The proposal originally planned for the complete sample of 58 3CR z $>$ 1 targets
and throughout Cycle 20 we obtained data for 22 of these 58. The observed sample
represent 38\% of our proposed sample. Of the 22 observed targets, 12 objects are
radio galaxies and 10 are QSOs. The observed sample spans a redshift range of $1.05 < z < 2.47$.

The names and properties of all the observed targets are listed in Table ~\ref{tab:objinfo}.

 In the case of 3C418, the source is at low galactic latitude and thus the field is contaminated by a large number of stars. The field is also heavily reddened \citep[A$_{V}$ = 2.9][]{hilbert16} making the analysis of the environment of this high-z object impossible. For this reason, we choose to exclude this target from the discussion of this paper. 

Both the UVIS (F606W) and IR (F140W) channels of WFC3 were used to image each of our targets.
The UVIS observations have a field of view of 162" x 162" with a pixel scale of 0.04". The IR observations cover a field of view of 123" x 136", corresponding to a projected distance of 1.0 Mpc x 1.2 Mpc at $ z = 1.5$. The pixel scale for WFC3 IR is 0.13".

We  first  download the  data  from  the  Mikulski Archive  for  Space
Telescopes (MAST).  We  customize the data reduction for  our data set
using two  different reductions for the  UVIS and IR images.    In
  the custom reduction for the UVIS  data, we first correct for charge
  transfer   efficiency  losses   using  the   algorithm  derived   by
  \citet{anderson10}, which produces the ``FLC'' (flat fielded and CTE
  corrected)  calibrated  files.   The   second  part  of  the  custom
  reduction focuses  on the  cosmic ray  removal through  a multi-step
  process.  To this  aim, we  use the  {\it Python}  version of  L.~A.
  Cosmic \citep{lacosmic} twice  on each image. The first  run is made
  with conservative  parameters, in order  to make sure that  only the
  obvious and  brighter cosmic rays  are removed, and no  real objects
  are affected.  A second L.~A. Cosmic run is then performed with more
  stringent parameters to remove the  cosmic rays in the region around
  the chip  gap only.  This  is needed since  we only have  two dither
  points, and therefore while the chip  gap is fully covered, there is
  a region of  the gap that is  only imaged once. At  least two images
  are needed  for the  {\it Astrodrizzle} cosmic  ray removal  task to
  work effectively.   {\it Astrodrizzle} is  then used to  combine the
  images,  and remove  residual cosmic  rays. However,  with only  two
  dither  points,  we  noticed  that not  all  events  are  completely
  removed. In particular,  pixels that are impacted by  cosmic rays in
  both images cannot  be corrected. In order to  remove these residual
  cosmic  rays we  first  make  a mask  that  includes pixels  showing
  significant flux  excess compared to the  surrounding pixels.  These
  are identified  by a  simple algorithm  that compares  each drizzled
  image with  both the difference  and the  ratio of the  original two
  images.   The marked  pixels  in the  mask are  then  grown using  a
  Gaussian kernel of  appropriate FWHM (generally $\sim$  1 pixel), in
  order to  fix a slightly larger  area. Pixels in that  area are then
  replaced by linear interpolation of the surrounding pixels using the
  {\it IRAF} taks {\it fixpix}.

The reduction  for  the  IR data  uses  the standard  HST
  pipeline followed by a persistence correction \citep{long11}. For more details regarding
  the steps of the custom reduction, see \cite{hilbert16}.

\begin{deluxetable*}{lrrcccl}
\tablecolumns{8}
\tabletypesize{\scriptsize}
\tablecaption{The observed sample.}
\tablewidth{0pt}
\tablehead{
\colhead{3CR Name} &    \colhead{RA} &        \colhead{Dec} &       \colhead{z} &           \colhead{$S_{178 MHz}$}             &       \colhead{Log $L_{178 MHz}$}         \\
                              &                       &                        &                       &    \colhead{(Jy)}                             &  \colhead{(erg/sec/Hz)}                        \\
                             }
\startdata
\cutinhead{Radio Galaxies}
3C 210      & 8:58:10.0  & +27:50:52             & 1.169        & 9.5           &  35.85         \\
3C 230      & 9:51:58.8  & -00:01:27            & 1.487         & 19.2          &  36.37       \\
3C 255      & 11:19:25.2 & -03:02:52          & 1.355   & 13.3          &  36.13    \\
3C 257      & 11:23:09.2 & +05:30.19         & 2.474    & 9.7   &  36.30  \\
3C 297      & 14:17:24.0 & -04:00:48             & 1.406    & 10.3\tablenotemark{e} &  36.05  \\
3C 300.1   & 14:28:31.3 & -01:24:08              & 1.159    & 10.1      &  35.87       \\
3C 305.1   & 14:47:09.5 & +76:56:22     & 1.132         & 4.6   &  35.50            \\
3C 322      & 15:35:01.2 & +55:36:53    & 1.168    & 10.2       &  36.19          \\
3C 324      & 15:49:48.9 & +21:25:38    & 1.206    & 13.6       &  36.04        \\
3C 326.1   & 15:56:10.1 & +20:04:20     & 1.825         & 9.0   &  36.19          \\
3C 356      & 17:24:19.0 & +50:57:40    & 1.079         & 11.3          &  35.85        \\
3C 454.1   & 22:50:32.9 & +71:29:19     & 1.841         & 10.2          &  36.25    \\
\cutinhead{QSOs}
3C 68.1     & 02:32:28.9 & +34:23:47    & 1.238    & 12.1       &  36.01   \\ 
3C 186      & 07:44:17.4 & +37:53:17    & 1.069    & 13.0       &  35.90     \\
3C 208      & 08:53:08.6 & +13:52:55    & 1.112    & 17.0       &  36.06   \\
3C 220.2   & 09:30:33.5 & +36:01:24     & 1.157    & 8.6                &  35.80     \\
3C 268.4   & 12:09:13.6 & +43:39:21     & 1.402    & 9.5         &  36.01    \\
3C 270.1   & 12:20:33.9 & +33:43:12     & 1.528    & 12.7       &  36.21    \\
3C 287      & 13:30:37.7 & +25:09:11    & 1.055    & 16.0       &  35.98        \\
3C 298      & 14:19:08.2 & +06:28:35    & 1.438    & 47.1       &  36.73  \\
3C 418      & 20:38:37.0 & +51:19:13    & 1.686         & 11.9          &  36.26         \\
3C 432      & 21:22:46.3 & +17:04:38    & 1.785    & 12.5       &  36.32        \\ 
\enddata
\label{tab:objinfo}
\tablecomments{In column 1 of the table we give the names of the objects. In columns 2 and 3 we display the coordinates. In column 4 we show redshifts. In column 5 we show the flux at 178 MHz , and in column 6 we show the logarithm of the luminosity.}
\end{deluxetable*}
\section{Photometry}
\label{photometry}
We first identify sources on the IR images and then we perform photometry on both IR and UVIS images based on the object catalog derived from the IR.
The emission of each galaxy in the UVIS images is dominated by
a younger stellar population since the rest-frame wavelength of the UVIS pass band (F606W) resides in the UV at $z >1$. These young stellar components usually appear as "blobby" structures (i.e. regions of star formation), and do not allow a straightforward identification of the object as a whole.  At 1 $< z < $ 2.5 the IR pass band (F140W) is rest frame optical and thus samples older stellar populations, resulting in a smooth regular shape.
For this reason, we use the IR images to identify sources.

The procedure is as follows. We use Source Extractor (SExtractor) \citep{bertin96} in the
MAG BEST mode for the IR images. Such a mode allows for measurements of the flux for sources with
different morphologies. When set in this mode, SExtarctor uses a flexible elliptical aperture
around every detected object and measures all of the flux inside that, provided that the
aperture is larger enough. SExtarctor constructs the specific elliptical apertures
using the Kron radius, described in \cite{kron80}. However, if the elliptical aperture is smaller than 3
pixels in radius, SExtractor defaults to a circular aperture of 3 pixels. Also, if the contribution form other sources
is determined to exceed 10\%, an isophote corrected flux/magnitude is used. This method retrieves the fraction of flux in the wings of an object that would be missed in the isophotal magnitudes by assuming a Gaussian profile. For full technical details see \cite{bertin96} section 7.4.2.

In order to select the optimal parameters we simulate a 512 x 512 pixel image convolved with a {\it Tinytim} \citep{krist11} model PSF
consisting of 78 galaxies representative of the galaxies within our targets' fields.
We run SExtractor on this image with varying parameters (Kron factor, and minimum radius) until we minimized
the difference between the known, simulated magnitudes and those given as output
by SExtractor. After establishing the ideal parameters, Kron factor of 2.5 pixels and minimum radius of 3.5 pixels, we run SExtractor on the IR images
to identify sources.

We use the SExtractor catalog produced from the IR data to
select the regions on which we perform aperture photometry on the UVIS images. As noted above since the galaxies in the UVIS
images are more likely to be irregularly shaped, we use aperture photometry to measure the flux of all the components
located within the region covered by the galaxy in the IR image.
The aperture size is adopted from the output of the SExtractor fit for the IR sources.
The specific radius used for each source is $r = \alpha \times R_{.9}$ where $R_{.9}$ is the 90\% effective light radius of the corresponding IR source.
After testing a range of values  on simulated galaxies of different  morphologies, we find that $\alpha = 1.2$ is
optimal to accurately measure the magnitude of the objects in the IR band. This value is also appropriate 
to encompass all of the flux of the individual components seen in the UVIS image that are co-spatial 
to a single source in the IR data.
After we complete the photometry we correct for galactic absorption and work with the AB magnitudes
(F606W zero point 26.0691 , F140W zero point 26.4524).

The final step in producing the source photometric catalog for each field is to remove objects identified as stars so that
the photometric data for each field contains only galaxies.

\section {Methods for Finding Clusters and Groups}
\label{methods}
We use two complementary
methods to establish which of our fields may contain a cluster or group. Each approach has limitations and so in order to
find the best candidates, we compare the results from both methods.

Firstly we look for the presence of a RS in CMDs in analogy with the so-called RS technique \citep{gladders_yee00}. This method is sensitive to massive clusters where the RS is more clearly defined. 
Secondly we identify over-densities of objects in the field of our targets when compared with the object densities from randomly selected fields. By comparing the results from multiple methods, we can better identify which fields are the best cluster candidates.
Below we describe the two different approaches.
\subsection{Red Sequences}
The existence of a RS in a CMD is a known indicator of clustering. A cluster's RS is identified as a linear relationship in CMD where bright (early-type) galaxies, dominated
by an old stellar population, are located. The position (color and slope) of the RS is determined by the evolution of the cluster early-type galaxies, and thus depends on redshift. Therefore, to determine the presence of a RS, we firstly identify the early-type objects
and determine whether their positions on the CMD coincide with the expected position of the RS at the redshift of the radio source in that field. 
\subsubsection{Morphological Classification}
\label{morphClass}
The first step in determining whether the CMD exhibits a cluster RS is to determine
which objects in the field are early-type galaxies (ETGs). We
morphologically classify all galaxies inside a 40'' (346 kpc at z = 1.5) radius encircled upon the target.

\begin{figure}
         \centering
        \includegraphics[height=1.7in]{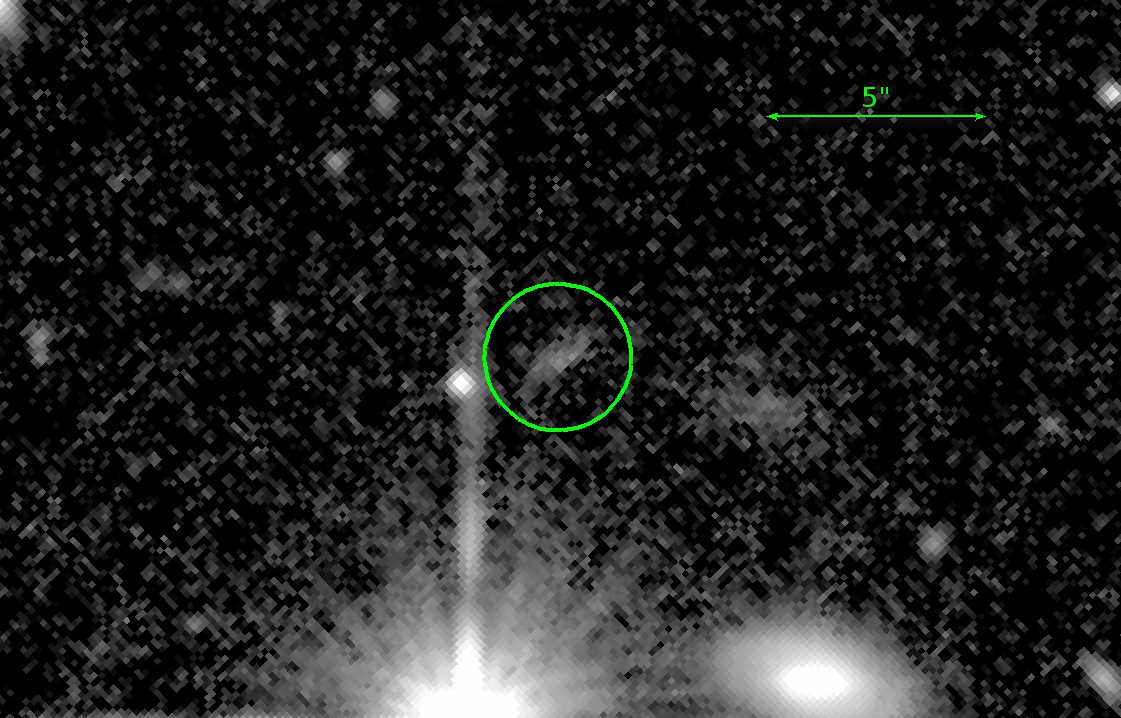}
        \caption{WFC3 IR F140W image of an object in the field of 3C68.1. This object was originally fit and estimated to have a S\`{e}rsic index within our early-type range.
         After visual inspection, it is clear that this object is in fact not an early-type galaxy and should not be marked as so in our CMD.}
         \label{rejected}
\end{figure}

To classify the galaxy morphologies, we fit the IR 2-D surface brightness profiles of all galaxies in each field using a S\`{e}rsic law in \textit{Galfit} \citep{peng10}.
The S\`{e}rsic law is given as
\begin{equation}
\Sigma(r) = \Sigma_{e} \left[-\kappa \left(\left(\frac{r}{r_{e}}\right)^{1/n} -1\right)\right],
\end{equation}
where $\Sigma_{e}$ is the surface brightness at the effective radius $r_{e}$ (half of the total flux is within $r_{e}$) and the S\`{e}rsic index is $n$.
We classify a galaxy as early-type if the best fit results in a S\`{e}rsic index, n, of $2 < n < 8$.
For each early-type galaxy we visually inspect the image as well as the residuals in order to determine if the fit was appropriate. In a small number of cases ($<2\%$) \textit{Galfit} misclassifies an object due to low signal to noise, or contamination from nearby objects. Fig.~\ref{rejected} shows an example of an object classified as early-type that we reject possibly due to low signal to noise or contamination. 

\subsubsection{Red Sequence Modeling}
\label{redsequencemodeling}
In order to produce a model red sequence estimate, we use {\it GalEv} \citep{kotulla}, the evolutionary
synthesis modeling program. The {\it GalEv} input parameters include mass, metallicity, and redshift of formation. For our models, we assume a single burst of star formation followed by passive evolution. Our goal is not to provide a detailed model of the evolution of single galaxies, but rather a comparison of different evolutionary states at the redshifts of our targets. Therefore a simplified framework, with a single star burst and passive evolution is sufficient. Furthermore, we tested that more complicated evolutionary models with non-instantaneous or multiple star forming events would not provide significant changes for the location of the predicted red sequences unless the star forming events are very recent.  {\it GalEv} outputs the modeled magnitudes starting from the redshift of formation. 

As a reference  we use the observed red sequence in the well-studied X-ray selected cluster RDCS 1252.9-2927 at z = 1.24 \citep{blakeslee03}. The cluster was observed with HST/ACS using the filters F775W and F850LP as part of program GTO/ACS 9290 and is known to have a well-defined RS even at such a high redshift.
We derive the parameters that produce the best representation of the observed RS for two different redshifts of formation (z$_1$ = 6.5, z$_2 $= 20). We model the evolutions of galaxies with two different masses ( i.e. corresponding to different magnitudes), which allow us to identify the slope in the CMD. The choice of parameters (mass and metallicity), are consistent with mass-metallicity relations described in \cite{lee08}. The derived set of parameters is then used to obtain the model magnitudes at each redshift for the two filters used in our WFC3 observations. The parameters used for the model are listed in Table ~\ref{GalParam}.

\begin{deluxetable}{lcc}
\tablecolumns{2}
\tabletypesize{\scriptsize}
\tablecaption{GalEv Parameters}
\tablewidth{0pt}
\tablehead{
\colhead{Mass ($M_\odot$)} & \colhead{} & \colhead{[Fe / H]}}
\startdata
\hline
\cutinhead{$z_{1} = 6.5$}
$1\times10^{10}$  & & 0 .0 \\
 $5\times10^{11}$ &  & +0.3 \\
\cutinhead{$z_{2} = 20$}
$1\times10^{10}$ & & -0.3  \\
$5\times10^{11}$ & & 0.0  \\
\enddata
\tablecomments{For the redshifts of formation (6.5 , 20), each row shows the GalEv parameters used to generate a galaxy used in our fit. Other parameters that remain consistent across all GalEv models are IMF: Salpeter IMF (.1 - 100 $M_\odot$), Burst: No Burst, Type: E (Elliptical), and Extinction Law: None.}
\label{GalParam}
\end{deluxetable}

\begin{figure*}[t!]
    \centering
          \centering
        \includegraphics[height=1.7in]{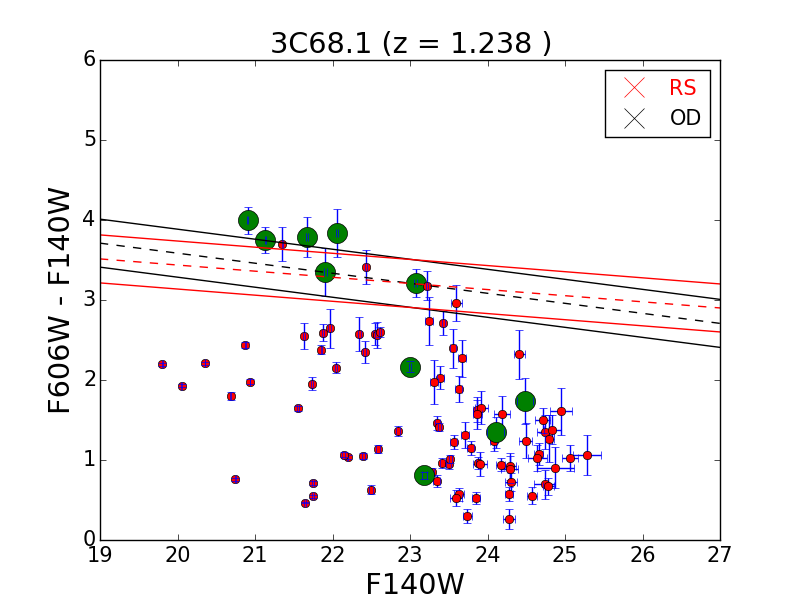}
    ~
        \centering
        \includegraphics[height=1.7in]{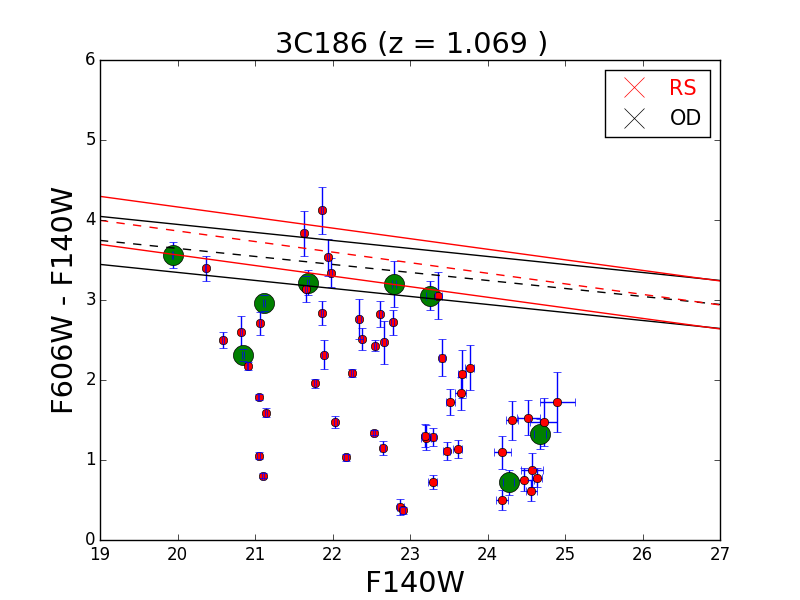}
    ~
        \centering
        \includegraphics[height=1.7in]{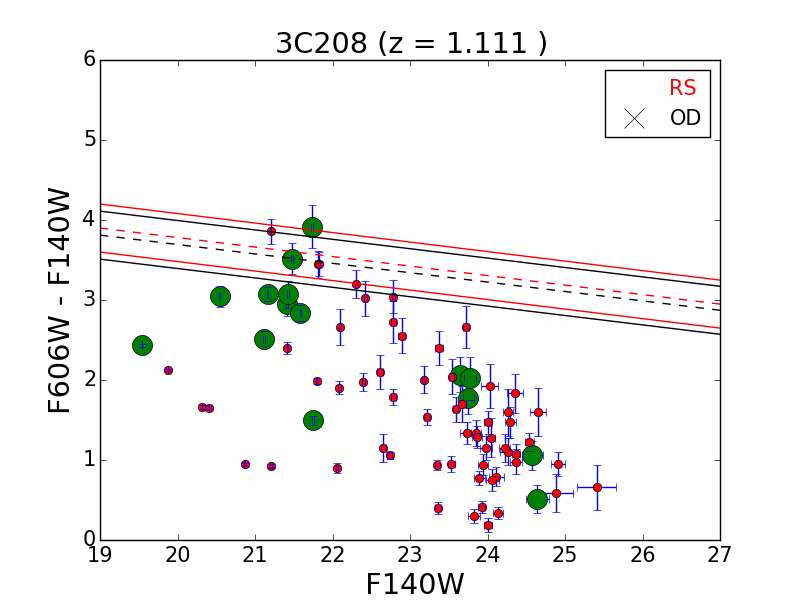}
    ~
        \centering
        \includegraphics[height=1.7in]{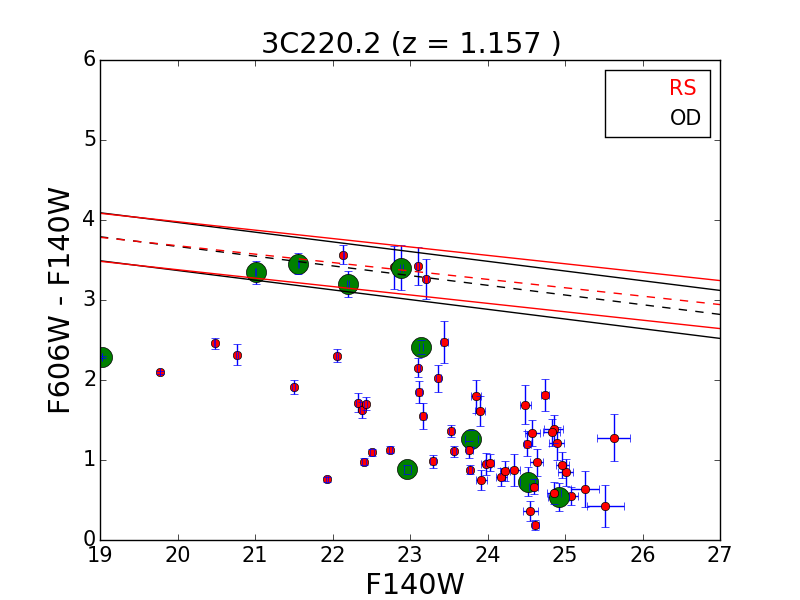}
     ~
        \centering
        \includegraphics[height=1.7in]{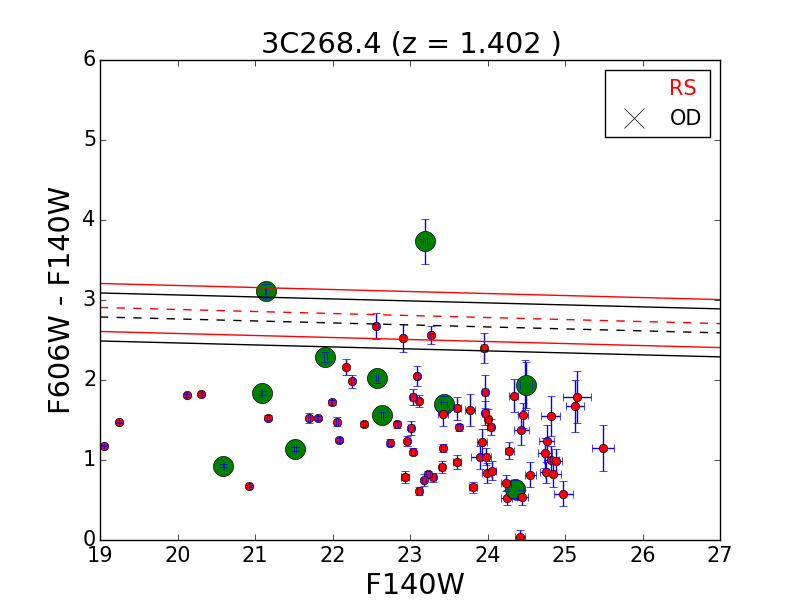}
    ~
        \centering
        \includegraphics[height=1.7in]{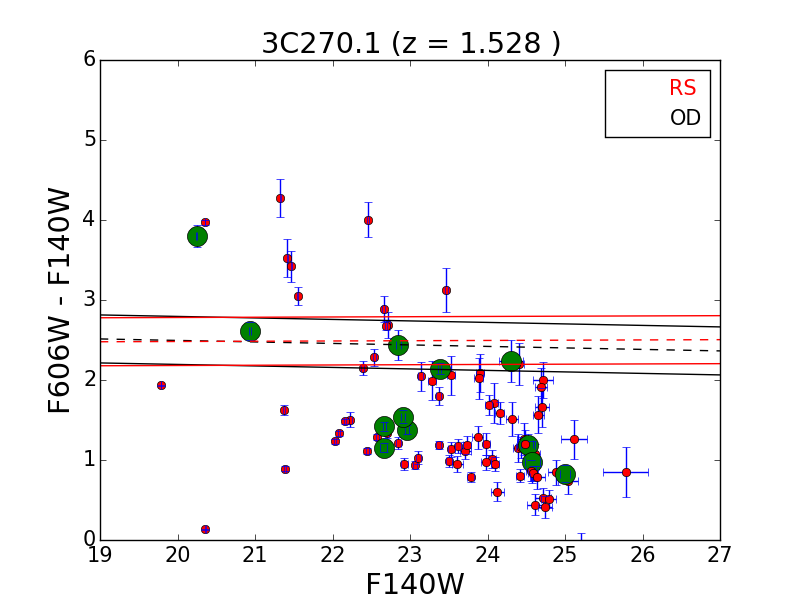}
 ~
        \centering
        \includegraphics[height=1.7in]{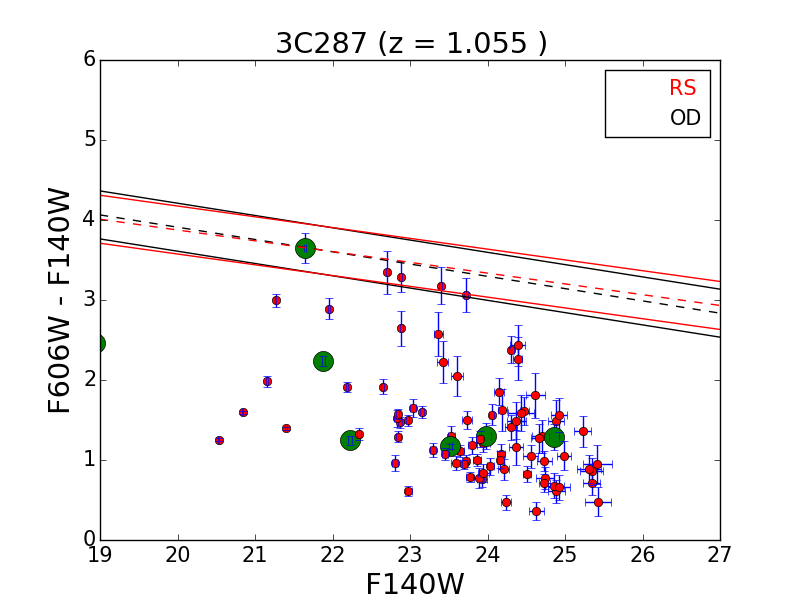}
 ~
        \centering
        \includegraphics[height=1.7in]{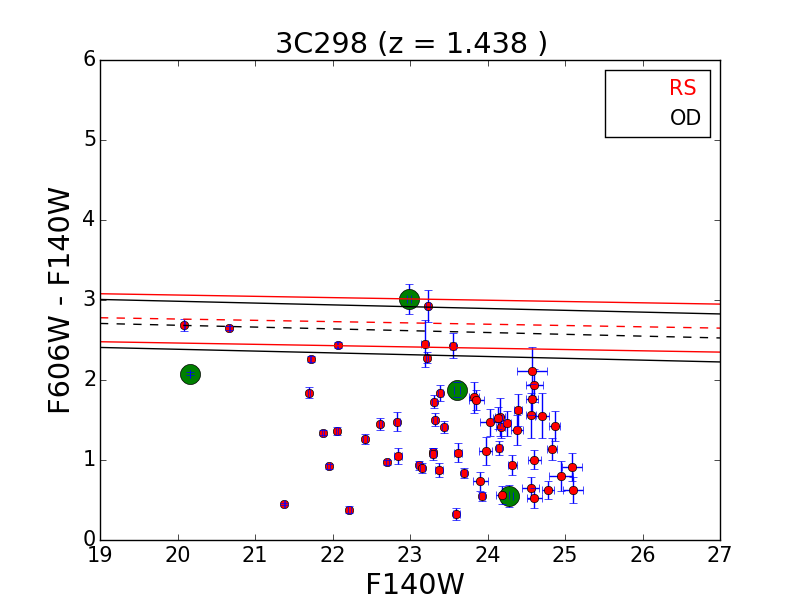}
 ~
        \centering
        \includegraphics[height=1.7in]{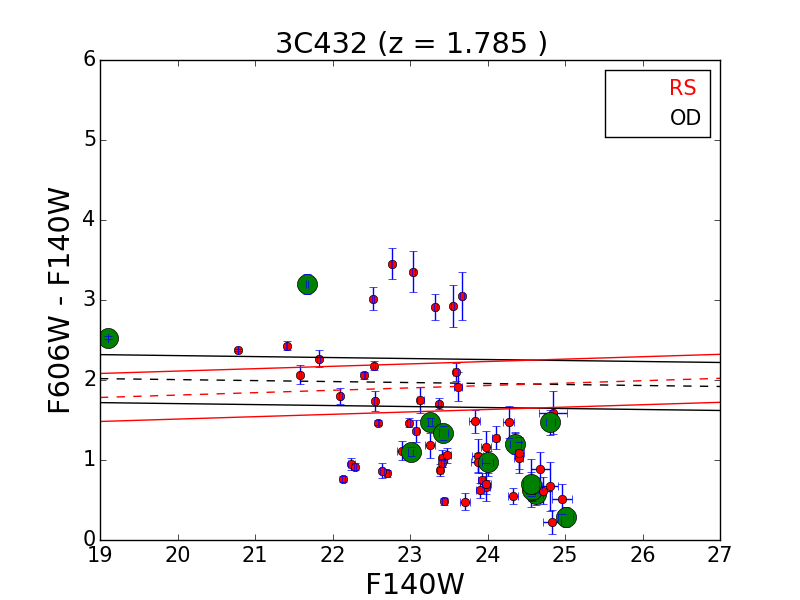}
    \caption{The CMD of the 9 QSOs in our sample. The plots contain all objects with magnitude less than 27 contained within a 40" radius of the target. The green circles, indicate that we classify the object as early-type, while blue circles represent objects that we originally classify as early-type but later reject the object due to contamination or another anomaly.  The red and black lines represent our model red sequences using GalEv parameters with a redshift of formation of 20 and 6.5 respectively. The dashed lines surrounding the models visualize a spread of $\pm$ 0.3 magnitudes. The target QSOs are not displayed in the figures. \label{CMD_QSO}}

\end{figure*}

\begin{figure*}[h!]
    \centering
        \centering
        \includegraphics[height=1.7in]{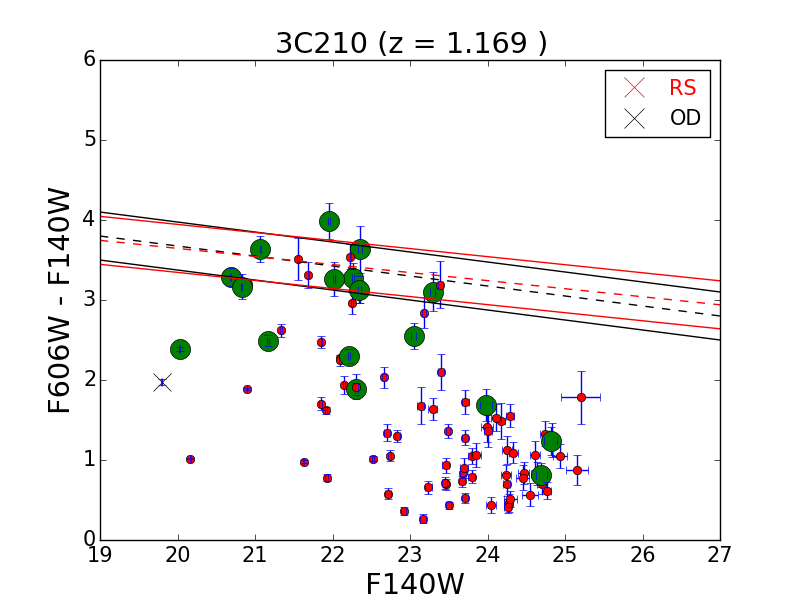}
    ~
        \centering
        \includegraphics[height=1.7in]{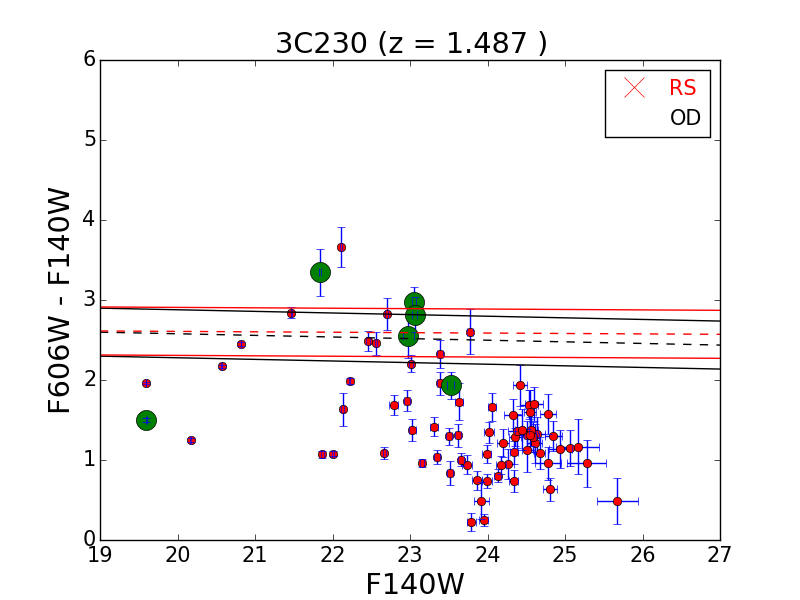}
    ~
        \centering
        \includegraphics[height=1.7in]{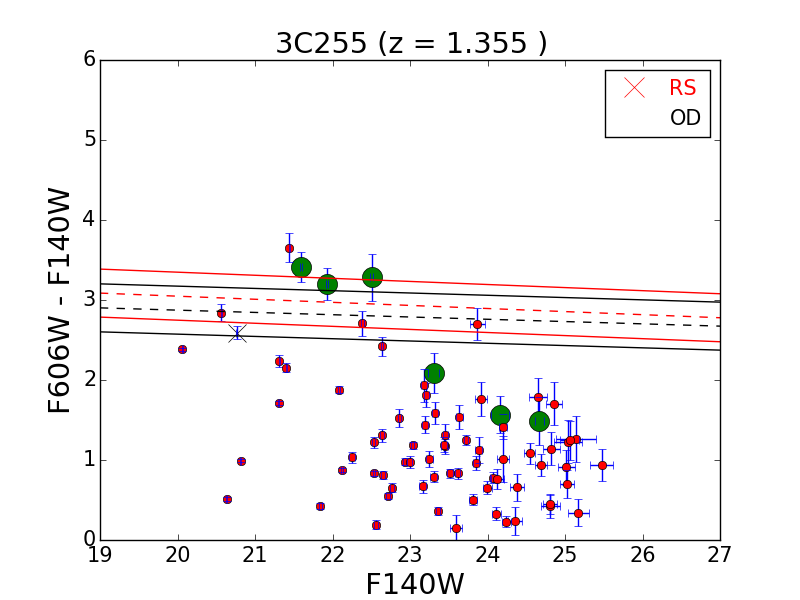}
     ~
        \centering
        \includegraphics[height=1.7in]{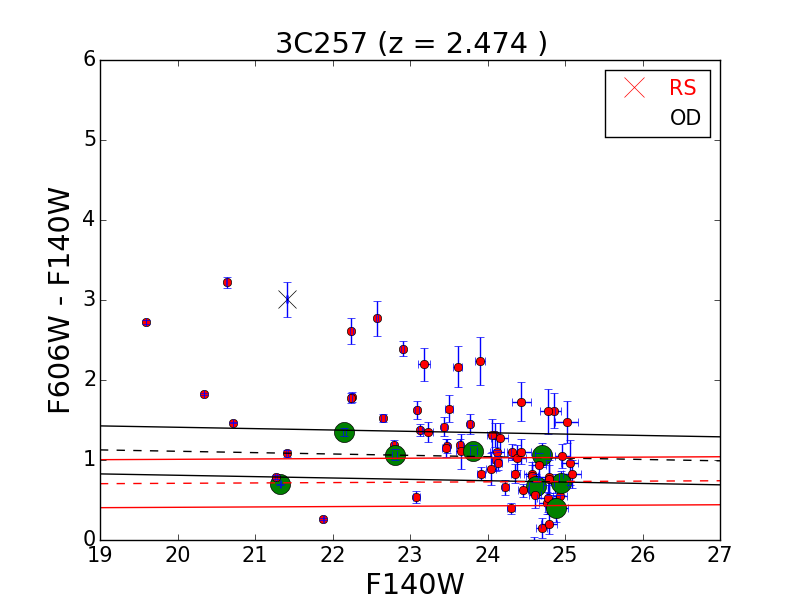}
    ~
        \centering
        \includegraphics[height=1.7in]{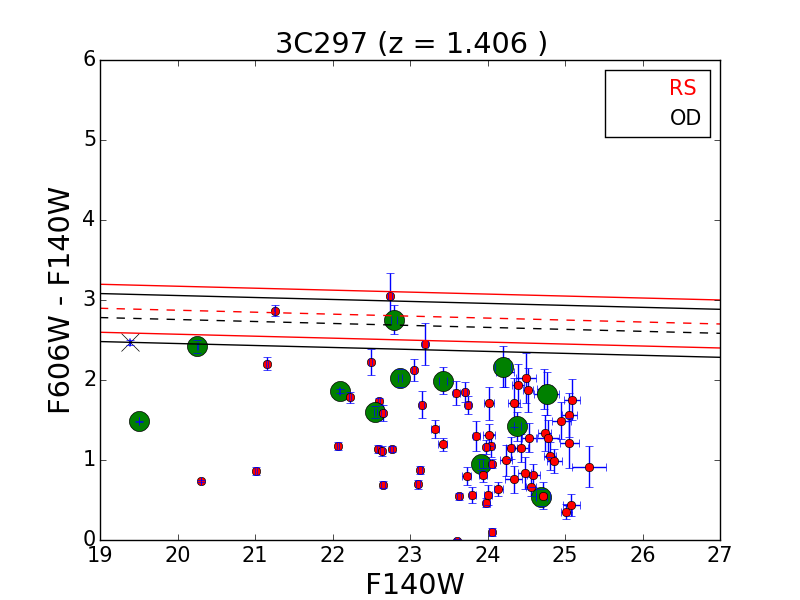}
     ~
        \centering
        \includegraphics[height=1.7in]{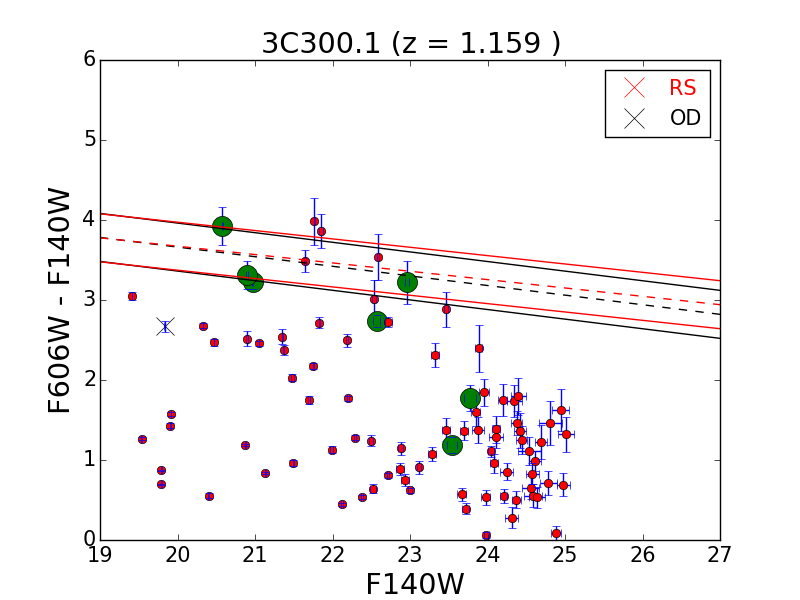}
    ~
        \centering
        \includegraphics[height=1.7in]{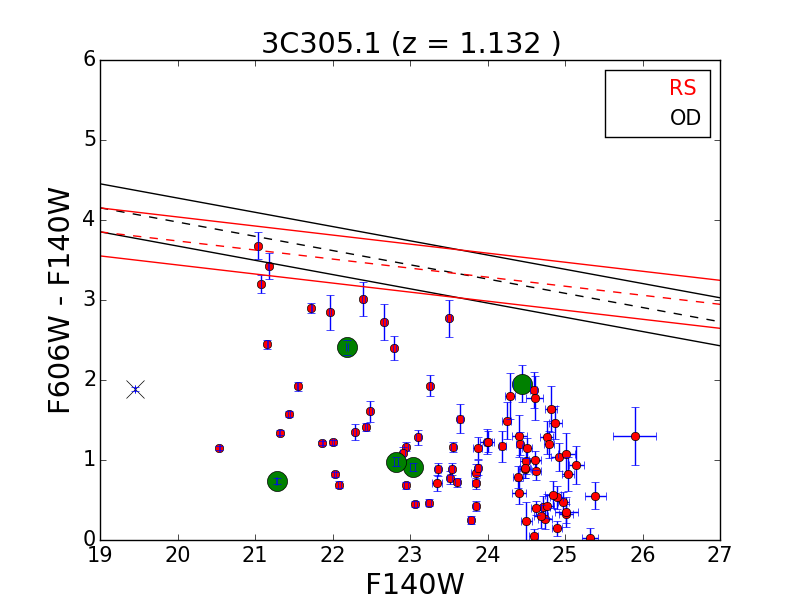}
 ~
        \centering
        \includegraphics[height=1.7in]{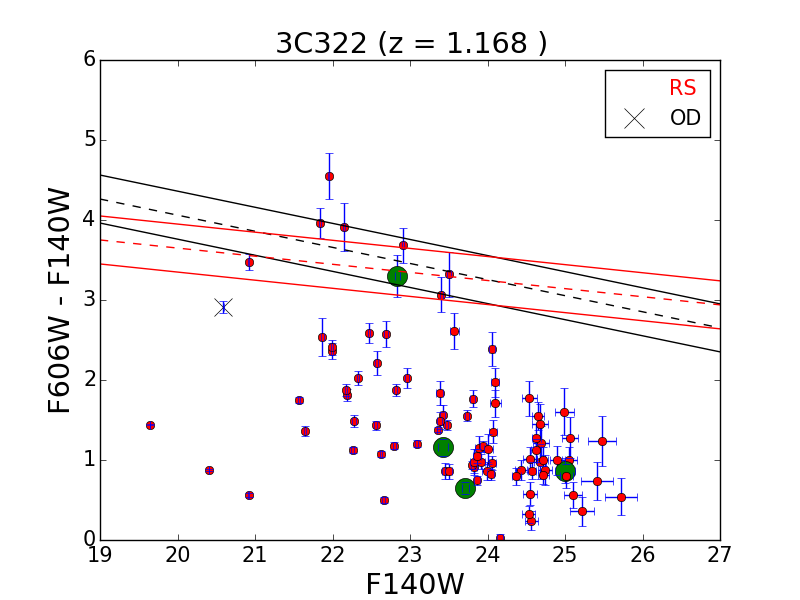}
 ~
        \centering
        \includegraphics[height=1.7in]{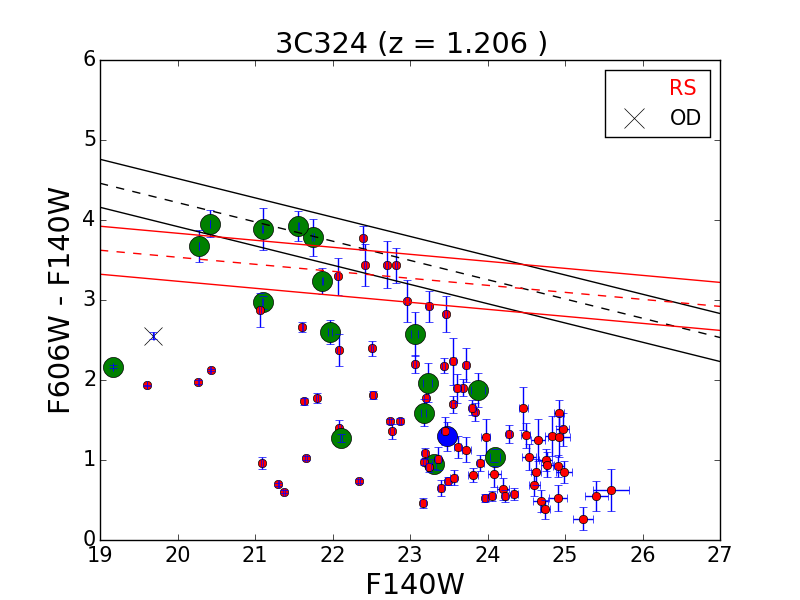}
 ~
        \centering
        \includegraphics[height=1.7in]{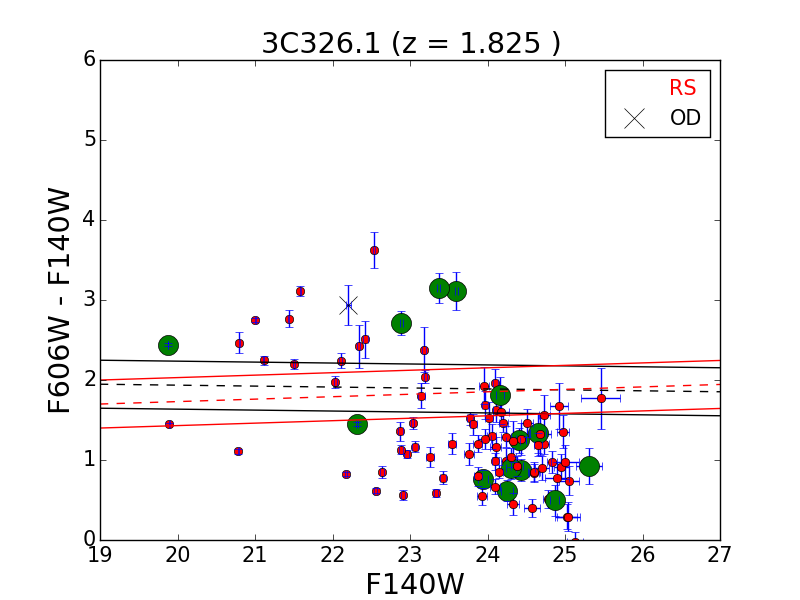}
 ~
        \centering
        \includegraphics[height=1.7in]{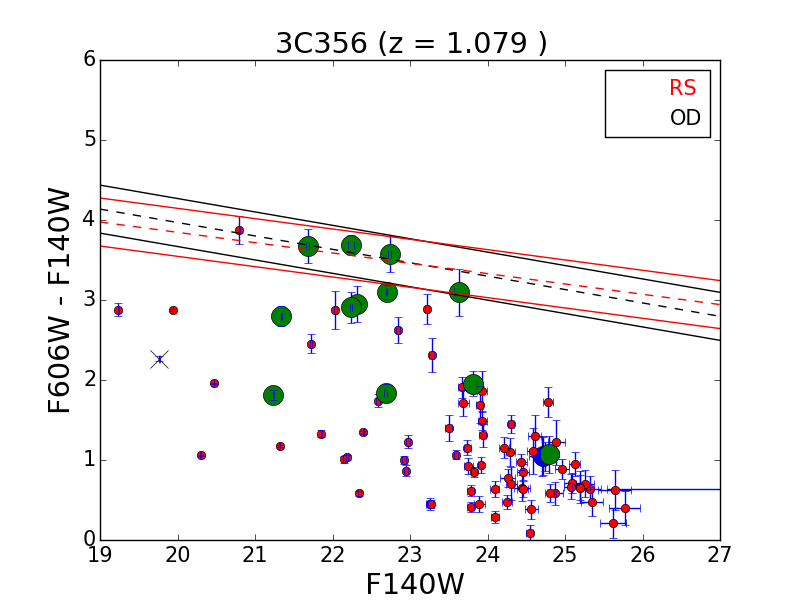}
    ~
        \centering
        \includegraphics[height=1.7in]{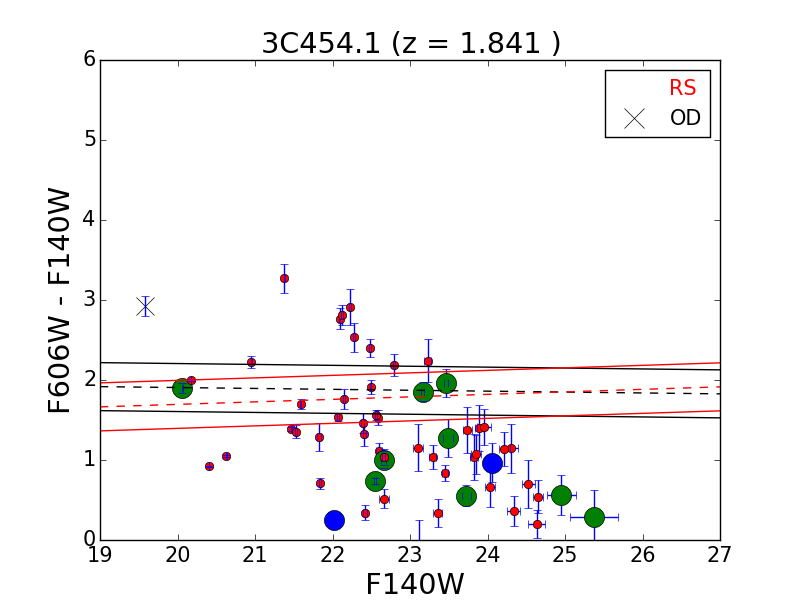}
    \caption{The CMD of the 12 RGs in our sample. The plots contain all objects with magnitude less than 27 contained within a 40" radius of the target. The green circles, indicate that we classify the object as early-type, while blue circles represent objects that were originally misclassified as early-type but later reject the object due to contamination or other anomaly (see section ~\ref{morphClass}).  The red and black lines represent our model red sequences using GalEv parameters with a redshift of formation of 20 and 6.5 respectively. The dashed lines surrounding the models visualize a spread of $\pm$ 0.3 magnitudes. A blue "X" represents the 3CR target.}
    \label{CMD_RG}
\end{figure*}

In Figs.~\ref{CMD_QSO}, ~\ref{CMD_RG} we show the resulting color (F606W $-$ F140W) plotted against the F140W
magnitude for each of the objects with $m_{F140W} < 27$ within a 40" radius surrounding the targets. Objects marked with green circles, indicate early-type galaxies (ETGs). Blue circles indicate that we originally classify the object as early-type but establish that the object was misclassified after visual inspection (as described in section~\ref{morphClass}).  In addition, we display the two model RSs corresponding to the two redshifts of formation ( z$_1$ in black, z$_2 $ in red). Given the uncertainties of the models and considering the range of redshift spanned by our sources, we highlight an area corresponding to a spread of $\pm 0.3$ mag where we qualitatively
expect to observe a RS.

Using the CMD, we define a cluster candidate as any field in which
we observe at least half of the galaxies classified as early-type lying within the area spanned by the models. We count all objects whose
1$\sigma$ error bar falls within the $\pm 0.3 mag$ band around either one of our two red sequence models. 

One of our targets (namely 3C~210) is known to reside in a well studied high-redshift cluster characterized by the presence of a red sequence \citep[]{stanford02}. 3C~186 is also known to reside in a cluster, but in that case the cluster was confirmed by the clear detection of X-ray emission from the intracluster medium \citep[]{siemiginowska10}. We use these clusters to test the reliability of our red sequence method. Reassuringly, this method correctly identifies both fields as cluster candidates because of the presence of a substantial number of ETGs within the region of the CMD where the corresponding model red sequences lie.

\subsubsection{Red Sequence Results}
Using this classification scheme we find 7 cluster candidates associated with 3 QSOs and 4 RGs. Results of this analysis are summarized in Table~\ref{countstable}. For each field, we report the number of objects falling within the 40" radius around the target ($n_{40"}$), the number of objects within the region that are classified as early-type ($n_{ETG}$), the number of ETGs whose 1$\sigma$ error bar falls within the $\pm 0.3$ color band around each of the RS models with redshift of formation 6.5, and 20 ($ETG_{1}$ and $ETG_{2}$ respectively). The objects classified as cluster candidates based the RS method described above are marked in column 6 with "RS".   

In addition to the 7 candidates identified with the RS method, we also point out that two additional fields (3C220.2 and 3C356, see Fig.~\ref{CMD_QSO} and Fig.~\ref{CMD_RG}, respectively) show a significant number of ETGs lying within the area spanned by the models. Our method does not identify these objects as cluster candidates because of the presence in the same field of a large number of bluer ETGs that do not fall close to the RS models. In 3C356 such ETGs are relatively bright, thus it is possible that at least some of them are foreground objects. Only spectroscopy of these objects can address this issue. 

Another object that shows an interesting population of blue ETGs is 3C257 (see Fig.~\ref{CMD_RG}), our highest redshift target ($z = 2.474$). Interestingly,  these blue ETGs lie exactly on top of the RS models. This will be further discussed in Sect.~\ref{BlueETG}.  

\subsection{Over-densities}
\subsubsection{Method}
In addition to the method using RSs described above, we investigate the existence of an over-density of galaxies in the regions surrounding the targets. The presence of over-densities could be an indication of clustering. Our method to search for significant over-densities compares the object counts in a region within 40" of each target against the average density of objects in control fields. 
 In the range of redshift between 1 and 2.5, the 
projected size corresponding to the radius we adopt changes by about 6\% (the smallest and largest size being approximately 347kpc at z=1.6 vs 326kpc at z=1, for the adopted cosmology). However, the main
concern is that the cluster core size might undergo a significant evolution between z=1 and z=2.5. Because of the poorly understood
relationship between cluster size and redshift, we prefer to keep the radius fixed, for the sake of simplicity.

The control fields are derived from a sample of 36 non-overlapping regions covered by the 3D-HST Survey data in the GOODS-S area \citep{brammer12}. Such a region was imaged using WFC3 IR and the F140W filter, i.e. the same configuration used in our 3CR observations. The selected regions in the 3D-HST data are chosen to avoid gaps present in the mosaic image. We create a catalog of objects in such regions using SExtractor. We manually remove any objects detected by SExtractor that are the result of artifacts.

\begin{figure}
         \centering
        \includegraphics[height=2.6 in]{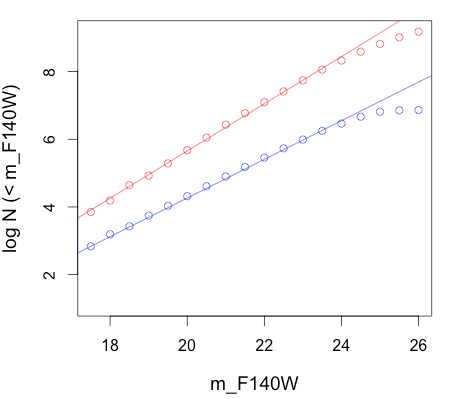}
        \caption{The $\log N - mag$ graph for our aggregated fields (blue dots) as well as the 3D-HST control sample (red dots). The y-axis is the logarithm of number of objects that have magnitudes less than or equal to the corresponding value on the x-axis. We fit the relationships for the region $19 < m_{F140W} < 24$ with linear models and display the resulting fits (red and blue for 3D-HST and 3C samples respectively). We shift the 3CR data downward by 2 on the $\log{N (< m)}$ axis (y-axis) in order to better display both sets of data that otherwise overlap.}
        \label{incomp}
\end{figure}

In order to ensure the completeness of the two samples, we select objects with $m_{F140W} < 24.5$ mag in both our 3CR fields, and the 3D-HST images. The upper bound of 24.5 magnitudes is derived from the $\log{N (< m)}$ vs. $m_{\rm F140W}$ plot for our sample as well as the comparison fields in the 3D-HST image. Fig.~\ref{incomp} is a modified version of the well known log N - Log S diagnostic, which allows us to determine at which flux (or magnitude) a survey becomes incomplete. In Fig.~\ref{incomp} the red dots represent the cumulative source distribution of the entire GOODS-S field covered by 3D-HST, while the blue dots are the data from our sample. In the figure we shift the 3CR data downward by 2 on the $\log{N (< m)}$ axis (y-axis) in order to better display both sets of data that otherwise overlap.

We see that significant deviations from the fitted lines in both cases occur for magnitudes fainter than 24.5. This is expected, since the exposure time of the 3D-HST pointings is only slightly longer than that of the 3C SNAPSHOT data. 

\subsubsection{Over-densities: Results}
\begin{figure}
        \centering
        \includegraphics[height= 3 in]{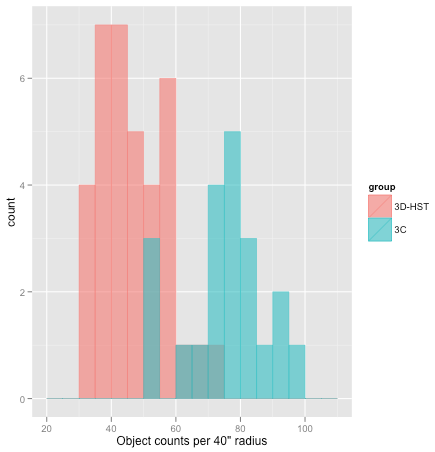}
        \caption{The distributions of number of objects within the 40" radius around the radio sources (blue), and the distribution of the number of objects in the randomly selected fields within the GOODS-S region (red).}
        \label{histograms}
\end{figure}

Firstly we test whether the radio loud AGN in our sample lie in over dense regions on average. Fig. ~\ref{histograms} presents the histograms for the distribution of counts amongst our sample (red) and the 3D-HST sample regions (blue). From visual inspection it is apparent that the number density of the objects in the 3CR fields is higher than in the control sample.
The mean of the object counts in the 3D-HST 40" radius regions is 45.9 with a standard deviation of 10.6 objects. The corresponding mean of the 3CR fields is 74.8 with a standard deviation of 13.2 objects. 
By comparing these two values we find that the environments of the radio sources are on average denser. This result comes from a student's t-test where we are able to reject the null hypothesis, i.e. that the two mean object densities are equal.  We determine that the mean of the object counts in the 3CR regions is higher than that of the control fields with very high statistical significance given by a p-value of $2.2\times10^{-10}$. The test was performed using the R function t.test in the stats package \citep{R}. 

In addition, we investigate the individual deviations of the number of objects in each of our fields from the average 3D-HST object density. We find that 4 out of 9 QSOs and 6 out of 12 RG environments show an over density with  $>3\sigma$ significance,  which corresponds to a p-value of $\sim 0.003$ for a normal distribution. In total, this amounts to $48\% \pm 20\%$ (the error corresponds to a 95\% Bayesian credible interval) of our sample of radio loud AGN being in over dense regions. The fraction of objects that lie in over dense regions for both the QSO and RG groups is highly uncertain due to the small sample sizes. The two fractions ($45\% ^{+28}_{-27}$ and $50\% ^{+24}_{-25}$ for the QSOs and RGs respectively) are statistically indistinguishable. 

Lastly, we perform a Bootstrap Kolmogorov-Smirnov test in order to rigorously test the difference in the distributions of the object densities in the fields of the QSOs and RGs. The result of such test is that we fail to reject the null hypothesis (p=0.904), i.e. that the distributions are different.

The objects classified as cluster candidates based upon the presence of an over density are marked in column 6 of Tab.~\ref{countstable} with "OD" (over density).   

Note that the density of galaxies in the central region of a cluster may depend on the dynamical state of the cluster itself, and that might affect our ability to identify cluster candidates. For example, if the cluster is in a merging state, its central regions may display a less pronounced distribution of objects compared to that of a relaxed cluster. Another possible complication is that the radio source might not be exactly at the center of the cluster, in particular for structures that are not relaxed yet. However,
for the specific goals of this work, we do not consider either of these scenarios  in detail.

As for the red sequence method in Sect.~\ref{redsequencemodeling}, we test our overdensity method against the two confirmed clusters in our sample, 
which are associated with 3C~186 and 3C~210. 
Our second method correctly identifies both fields as cluster candidates because of the presence of a significant overdensity of objects around the radio source.

\begin{deluxetable}{lcrrrcrl}
\tabletypesize{\scriptsize}
\tablecaption{Summary of Classification Methods}
\tablewidth{0pt}
\tablehead{
\colhead{3CR Name}&\colhead{$n_{40"}$} & \colhead{$n_{ETG}$} & \colhead{$ETG_{1}$} &\colhead{$ETG_{2}$}& \colhead{Cluster Candidate}}
\startdata
\hline
\cutinhead{QSO}
3C68.1         & 90 & 9 & 5 & 5 & RS, OD   \\
3C186          & 78 & 8 & 4 & 4 &  RS, OD \\
3C208          & 89 & 15 & 3 & 2 & OD  \\
3C220.2       & 50 & 10 & 3 & 4 &  No \\
3C268.4       & 79 & 11 & 1 & 1 &  OD  \\
3C270.1       & 75 & 12 & 4 & 4 &  No  \\
3C287          & 70 & 7 & 1 & 1 &  No \\
3C298          & 52 & 4 & 1 & 1 & No \\
3C432          & 65 & 12 & 0 & 0 &  No  \\
\cutinhead{Radio Galaxies}
3C210         & 90 & 17 & 9 & 9 &  RS, OD  \\
3C230         & 74 & 6 & 3 & 3 &  RS  \\
3C255         & 77 & 6 & 2 & 3 &  RS  \\
3C257         & 50 & 8 & 7 & 7 &  RS  \\
3C297         & 64 & 12 & 2 & 1 &  No  \\
3C300.1      & 99 & 7 & 4 & 4 &  RS, OD \\
3C305.1      & 70 & 5 & 0 & 0 &  No \\
3C322         & 81 & 4 & 1 & 1 &  OD  \\
3C324         & 81 & 15 & 5 & 5 &  OD  \\
3C326.1      & 79 & 14 & 1 & 1 &  OD \\
3C356         & 73 & 12 & 5 & 5 &  No  \\
3C454.1      & 84 & 9 & 3 & 4 &  OD

\enddata
\tablecomments{The first column shows the name of the 3CR target for a particular field. In column 2 we report the number of objects falling within the 40" radius around the target ($n_{40"}$). In column 3 we show the number of objects within the region that are classified as early-type ($n_{ETG}$). Columns 4 and 5 refer to the number of ETG whose
1$\sigma$ error bar falls within the $\pm 0.3$ magnitude band around each of the RS models with redshift of formation 6.5 and 20 ($ETG_{1}$ and $ETG_{2}$ respectively). In column 6 we indicate with  "RS" and/or "OD" whether the method based on the red sequences or the over densities, respectively, classifies the field as a cluster candidate.}
\label{countstable}
\end{deluxetable}

\begin{figure*}
\centering
   \includegraphics[width=18cm]{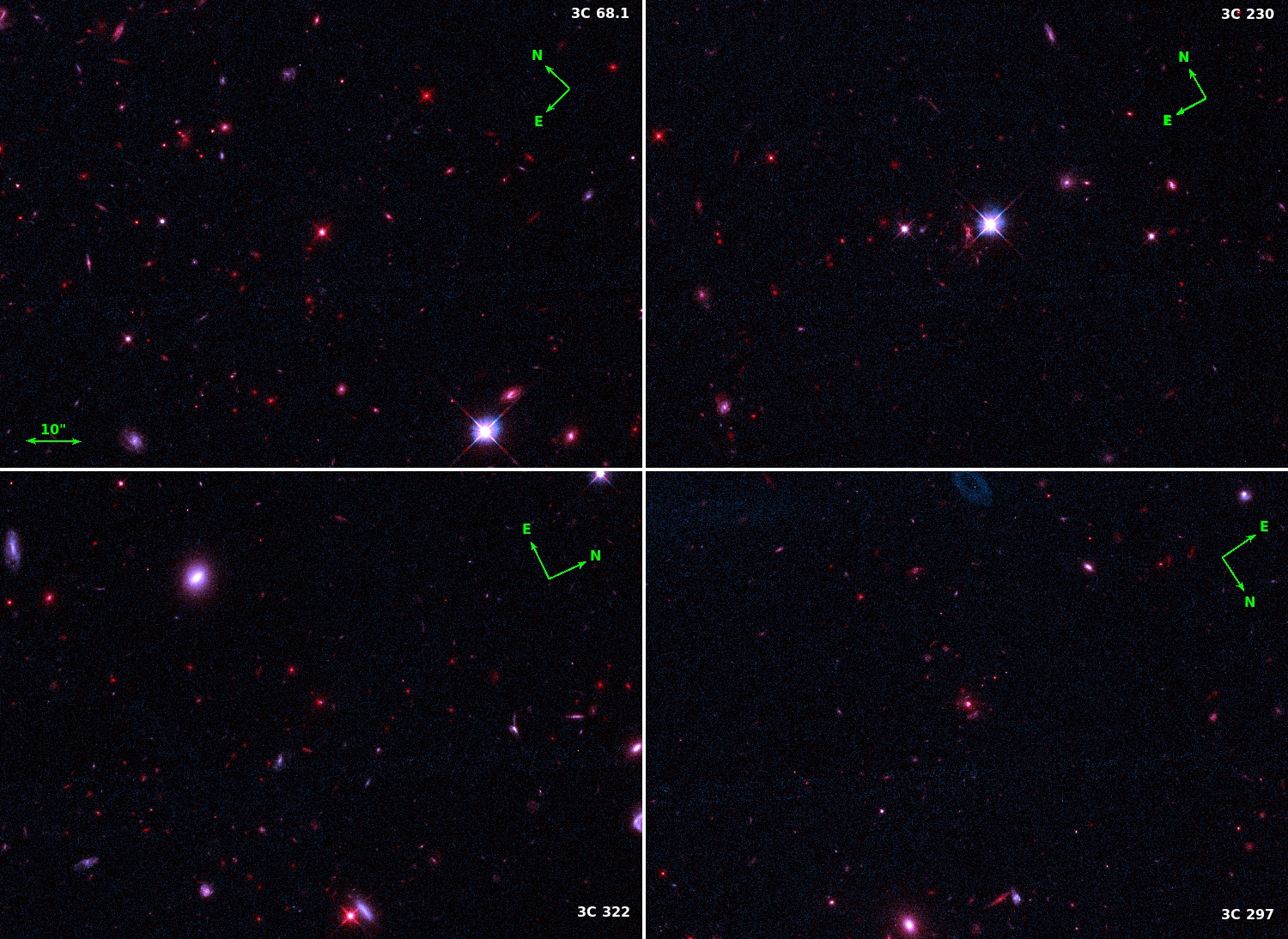}
   \caption{Four fields of 3CR RL AGN observed with HST WFC3 as part of program GO13023. The rgb images are made using the WFC3 F140W filter for the R channel, and the WFC3 UVIS F606W filter for both the G and B channels. the size of each image roughly corresponds to the field of view of WFC3 IR. The 10" reference corresponds to 84.6 Kpc, 86.0 Kpc, 83.8 Kpc, and 85.7 Kpc for 3C68.1, 3C230, 3C322, and 3C297, respectively. In the figure we show one case from each outcome of the cluster classification scenarios. In the upper left panel the field of 3C68.1 is classified as a cluster on the basis of the presence of both a RS and an over density. The field of 3C230 shows a RS but no over density. The field of 3C322 shows an over density but no RS. Lastly, the field of 3C297 shows neither a RS nor an over density. See Section~\ref{discussion} for details.\label{four_fields}}
\end{figure*}

\section{Discussion}
\label{discussion}

We used two different methods to investigate the environment of 3CR radio galaxies and QSOs at $z > 1$. The method based on the object over densities returns a larger number of cluster candidates (11) as compared to the method based on the presence of a RS (7). 

In four cases (2 QSOs and 2 RGs, namely 3C68.1, 3C186, 3C210, 3C300.1) both methods return a positive result.  
As already discussed above, two of these objects are already known to reside in clusters comfirmed by either the detection of  X-ray emission 
from the intracluster medium \citep[3C~186,][]{siemiginowska10} or by the presence of a well established red sequence \citep[3C~210,][]{stanford02}. 
These two clusters are identified correctly by both methods, i.e. we 
find both an overdensity of sources and a substantial number of ETGs lying on the region of the CMD where the red sequence is expected at the redshift 
of each of the two targets. 
The fact that those two confirmed clusters are correctly identified by both of our methods
gives us confidence that our cluster candidates are robust at least for the cases in which both of our methods agree on a positive detection.
However, all of our cluster candidates must be confirmed by other means, e.g., with spectroscopic information on the redshift of each
cluster galaxy member.

For the remaining two objects for which both of our methods return a positive result, there is no information in the literature about any association with a cluster of galaxies, to the best of our knowledge. However, we point out that the presence of a RS is very clear from their CMDs in Figs.~\ref{CMD_QSO} and ~\ref{CMD_RG}. This is particularly evident in the case of 3C68.1 in which 5 out of the 9 ETGs lie close to the RS models. 

Three RGs (3C230, 3C255, 3C257) are identified as cluster candidates by the RS method but they do not show an over density of objects. Even if at least half of the ETGs fall within the area spanned by the models, a RS is not clearly identified in 3C230, since they show no clear linear relationship. In 3C255 only 5 objects appear to define a RS. 3C257 is a peculiar case that we discuss in section~\ref{BlueETG}. It is possible that these 3 radio sources reside in small groups that include a small population of ETGs, but the evidence for the presence of a cluster is not convincing.

Six objects, two QSOs (3C208, 3C268.4) and four RGs (3C322, 3C324, 3C326.1, 3C454.1) show significant over densities but no evidence for a RS. In the case of 3C322 (Fig.~\ref{CMD_RG}) only one of the red objects that lie on the RS models is determined to be an ETG.  Conversely, in the field of 3C324 we see a relatively large number of red ETGs falling onto the models. However, this population does not represent the majority of the ETGs in the field. Thus our method does not identify this as a cluster candidate. 

The remaining 5 QSOs and 3 RGs are not identified as cluster candidates according to either of our methods. 

In Fig.~\ref{four_fields} we show four example fields, one for each category. In the top left panel the field of 3C68.1 is classified as a cluster by both the RS and the over density methods.   
The presence of a large number of red objects in this dense field is apparent from visual inspection. 
In the top right panel, the field of 3C230 is classified as a cluster according to a RS method, but there is not a significant over density of objects.
In this case, it is possible that a cluster is present, but since the object is located at $z \sim 1.5$ a significant number of the cluster galaxies are below our detection threshold. 
Alternatively, this object could be located in a group that includes a smaller number of galaxies with respect to rich clusters.
In the lower left panel we display an example of a field (surrounding 3C322) that shows a significant over density, but no RS. 
Note that despite the presence of a large number of red objects, they are not identified as ETGs, and thus the RS method fails to classify this field as a cluster.
In the lower right panel we show the field of 3C297, in which neither the RS method nor over density method classifies this field as a cluster. 
The paucity of objects in this field is clear from the image. 

We do not find any evidence for a statistically significant correlation between the number of 
galaxies detected within the 40'' radius and redshift, or between the number of ETGs and redshift. However, 
there are only four sources in our observed sample at z$>1.6$. Therefore it is difficult to identify any possible trends with redshift. It is interesting to note that the most distant object 3C~257 lies at the lower end of the galaxy number count
distribution, but such an object is not the one for which the 40" radius corresponds to the smallest projected 
size. It is possible, in that specific case, that the surface brightness detection limit of our observations 
may be playing a role.  Alternatively,
the smaller number count might be due to the younger age of the cluster, which may correspond to a significantly larger core radius.
But a much larger number of objects should be observed in order to statistically test these possible scenarios.

As the statistical analysis shows, a fraction consistent with about one half of our targets are associated with cluster or group candidates. This is in agreement with previous results using ground-based observations. In fact,  \cite{hillLilly91} found that about $50\%$ of powerful 3C radio galaxies at $z \sim 0.5$ inhabit rich clusters. Similar results were also found by \cite{best00} based on ground-based near IR observations. Two of our objects (3C324, 3C356) are in common with the sample presented in that work. Qualitatively, we conclude that our results are consistent with the published CMDs in that work. 

\subsection{Blue ETGs in the field of 3C257}
\label{BlueETG}
\begin{figure*}
   \plotone{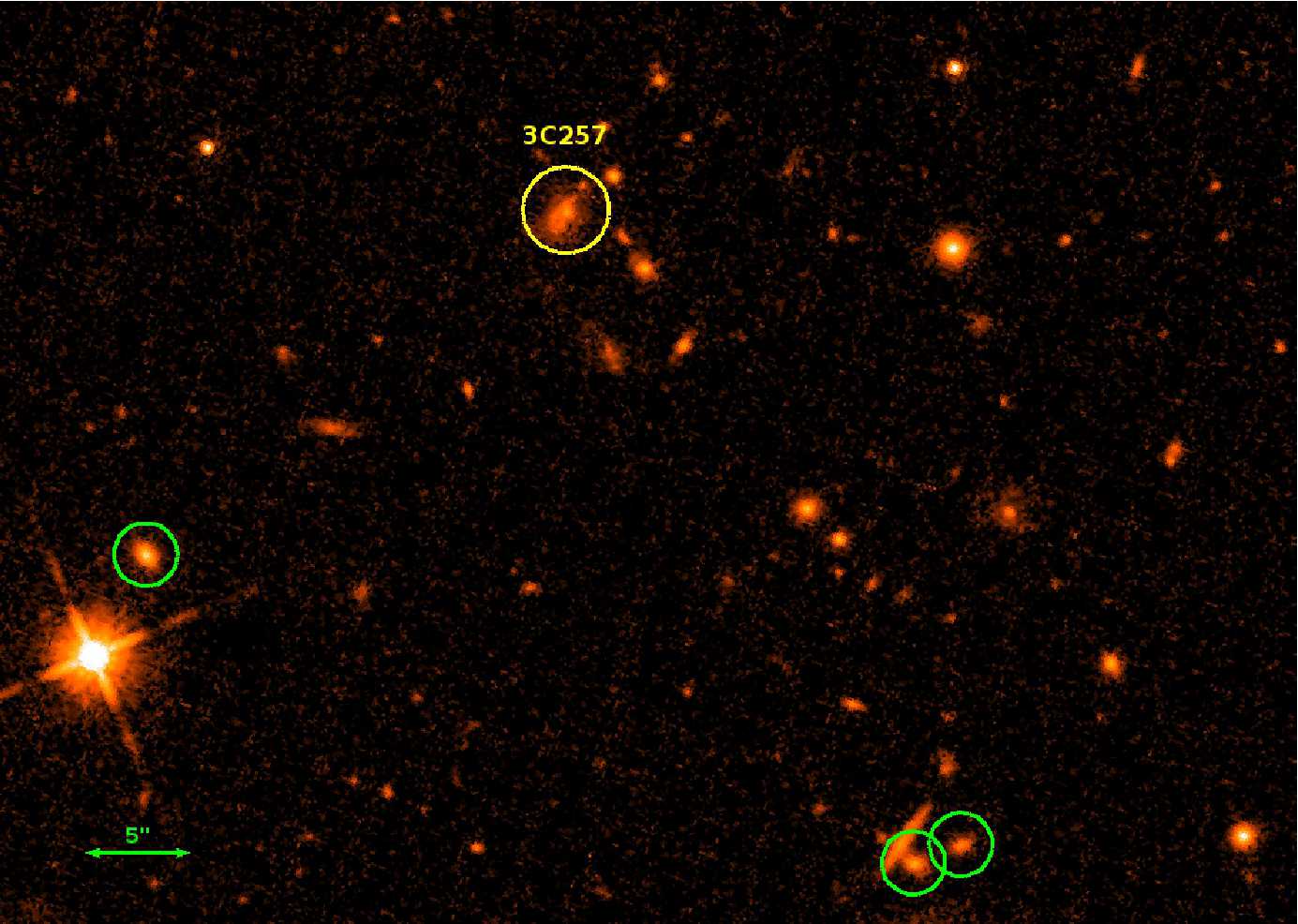}
    \caption{HST WFC3 IR image of a $60" \times 40"$ region surrounding 3C257. The 5" reference corresponds to 40.5 Kpc at the redshift of the target. The yellow circle displays the target (3C257) and the green circles show the three brightest ETGs in the field.}
    \label{blueETG}
\end{figure*}

\begin{figure*}
    \centering
        \centering
        \includegraphics[height=3in]{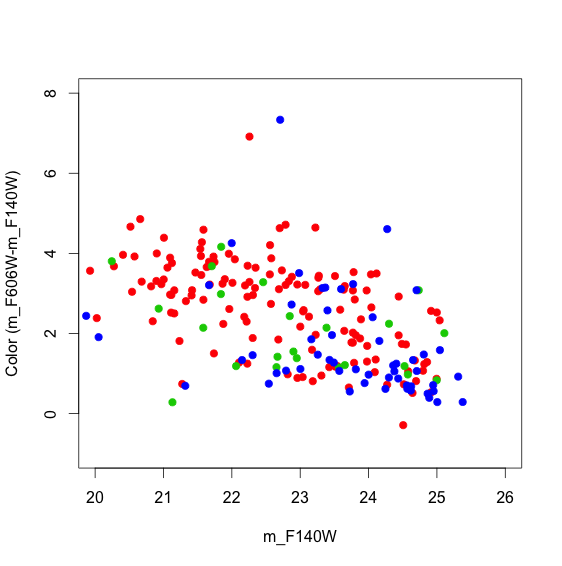}
    ~
        \centering
        \includegraphics[height=3in]{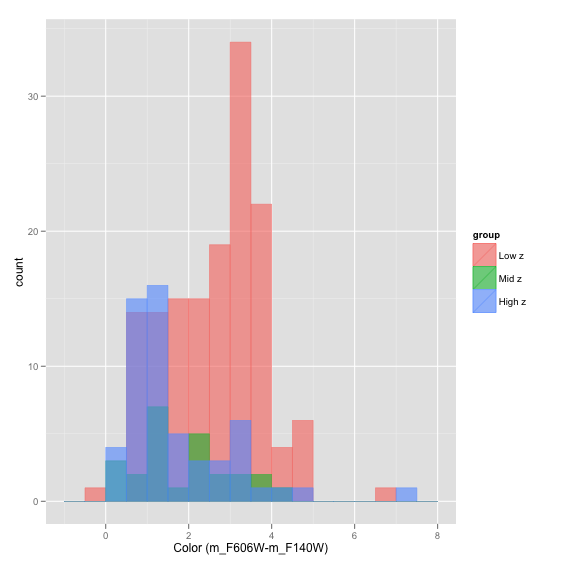}
 \caption{The left panel shows all of the ETGs identified in all of the 3CR fields. Different colors are assigned to different redshift bins. Red, green and blue represent $z<1.3$, $1.3<z<1.75$ and $z>1.75$. Note that the redshifts correspond to the measured redshift of the radio loud AGN at the center of each of the fields in which the ETG is located. The right panel shows the distribution of the color of the ETGs for each of the redshift bins previously specified.}
    \label{kotyla_seq}
\end{figure*}

In the case of the most distant source, 3C257 (z = 2.474), we observe a particularly interesting feature.
The derived RS models for the targets redshift predict that the ETGs should display an F606W-F140W color $\sim 1$, i.e. they should fall in the "blue" region of the CMD.
This is straightforward consequence of the fact that the models are reproducing the evolution of the stellar populations included in these galaxies since the first burst of star formation. 
At redshifts above $\sim 2$ a significant fraction of stars are still blue even in passive evolving ellipticals. Strikingly, a number of relatively bright ETGs in the field
of 3C257 are indeed blue and fall exactly in the region spanned by the models. In Fig.~\ref{blueETG} we show the WFC3 IR image of a region of $60" \times 40"$ in the vicinity of our target. 3C257 is marked with a yellow circle and the three brightest ETGs are marked with green circles. These objects lie at a projected distance of 200-300 kpc from the source. These galaxies may be similar to the star forming blue ETGs associated with one proto-cluster and a group recently found by \citet{mei15} in the Hubble Ultra Deep Field.

Interestingly, a population of redder (F606W - F140W $\sim 3$) objects is present in
the field of this galaxy. However, these objects are not identified as ETGs. Visual inspection reveals that some of these galaxies, including the host galaxy of 3C257, are close companions of target (Fig.~\ref{blueETG}). These objects display disturbed morphologies that is most likely indicating an active merger phase. We argue that these galaxies are reddened by the presence of a significant amount of dust. They are not recognized as ETGs because their morphology is very irregular. In the figure, these are the objects within about 10$"$ from the target. Other red galaxies are located further away from the target and their association with the RG is less obvious. This hypothesis may be further tested using multi wavelength observations e.g. in the infrared or at radio wavelengths with ALMA. However, there is already strong evidence from Herschel observations that at least for the host galaxy of the radio source, extremely high star formation activity is present \citep{podigachoski15}.

Without spectroscopic information we cannot exclude that the blue ETGs in this field are a population of lower redshift galaxies. We are convinced that these objects are not foreground ETGs for two reasons: i) a population of relatively bright blue ETGs is not commonly present in our sample; ii) their location is consistent with the prediction of the RS models. We speculate that this population of blue objects could be galaxies that will evolve into the RS galaxies observed in lower redshift clusters \citep{kodama98,kodama07}.

Curiously, 3C257 is one of the very few fields in our sample in which we do not detect a significant over density of sources.
Since surface brightness depends on redshift as $(1+z)^{-4}$, this may result from incompleteness due to the detection threshold of our data.
Another possible scenario that may explain the lack of a significant over density in this field is that  at $z>2$ proto-clusters are significantly more extended than the projected region of the sky covered by our observations, and less dense than similar structures at lower redshifts.

The presence of a trend between the color of the ETGs and redshift seems to be confirmed by the results shown in Fig.~\ref{kotyla_seq}. In the left panel we show the CMD for all of the ETGs identified in all of our 3C fields. Different colors identifies the redshift range of the radio galaxy in which each ETG is found. Red is for $z<1.3$, green for $1.3<z<1.75$, and blue is for $z>1.75$. The lower redshift ETGs are clustered around a color $m_{F606W}-m_{F140W} \sim 3$. The higher redshift ETGs are instead around $\sim 1$. The intermediate objects are scattered of a large range of colors. This is even more clearly visible in the right panel of Fig.~\ref{kotyla_seq}, which shows histograms for the color distribution of the ETGs. This may be evidence that the ETGs in the fields of radio galaxies between z=1 and 2.5 undergo dramatic evolution from star forming to passive evolving red sequence galaxies. 

 \subsection{Location of  3CR Targets in the color-magnitude diagrams.}
The location of the hosts of the 3CR radio galaxies in the CMDs with respect to the areas spanned by the RS models differs across the sample. We find that most of the radio galaxies have a color (F606W-F140W) $\sim$ 2-3 mag. The evolutionary models for red sequence ETGs predict bluer colors as redshift increases. For the lowest redshift radio galaxies ($z < 1.3$), the hosts are bluer than the RS models for the corresponding redshift. In the case of 3C255 and 3C297, with redshifts $z = 1.35$ and $z = 1.41$ respectively, we find that the host falls within the area spanned by the models. In the highest redshift RG in the sample, 3C326.1, 3C454.1, and 3C257 ($z = 1.83$, $z = 1.84$, and $z = 2.47$ respectively) the hosts' positions lie above the models in the CMD. The fact that the colors of the radio galaxy hosts do not follow the predictions of simple passive evolution directly implies that modeling the evolution of these objects requires additional parameters. For example, it is possible a major role in defining the stellar population is played by galaxy mergers, which appear to be closely connected to radio loud AGN activity \citep{chiaberge15}. Therefore the combined presence of dust and enhanced star formation associated with those events may be responsible for the observed color of each object at different redshifts. Detailed modeling of these effects is beyond the scope of this work, and will be performed in a forthcoming paper. 

\section{Conclusions}
\label{conclusion}

We studied the environments of 21 3CR radio loud AGN in the redshift range $ 1.05< z <2.47$ using HST Snapshot observations taken with WFC3 UVIS and IR at rest frame UV and optical wavelengths, respectively. 
We derived color vs. magnitude diagrams for all of the fields in order to search for red sequences defined by early-type galaxies at the corresponding redshift of each radio source. We find that for seven targets the majority of the ETGs are located within the area spanned by red sequence passive evolution models.  
The second method classifies fields around the targets as being over-dense if the object density is significantly higher than that of a control sample taken from 3D-HST GOODS-S. We determined that 3CR radio loud AGN at $z>1$ are associated with over-dense fields, on average. We also compare single targets with the average density of the control fields and we conclude that in 10 out of 21 cases the fields are over dense at more than 3$\sigma$. 
Our results are consistent with a fraction of $\sim 50\%$ of objects being associated with clusters or groups, in agreement with previous results \citep[e.g.][]{hillLilly91,best00}.

For a number of objects the two methods do not agree on the cluster classification. In the cases where only the RS method classifies the field as a cluster, one possible explanation may be that the target resides in a lower density cluster or a group that includes a population of passively evolving ETGs. Alternatively, and particularly for the higher redshift objects, we may be missing a substantial number of faint cluster galaxies because of the inherent detection threshold of our data. Since most of our targets are clustered at $z<1.5$, it is difficult to statistically test if such an hypothesis is viable using our data. A larger number of high-z 3C fields must be observed to allow for proper statistical analysis. This would allow us to both firmly establish whether the observed number density of galaxies around each radio source depends on redshift, and understand whether that is due to the detection limit of the survey. Alternatively, the lower number density of objects around more distant targets might be a result of the different dynamical status of those clusters.

For the opposite case in which an over density is present but there is no RS, the most likely explanation is that evolution is playing a role. A significant population of red objects are usually present but they are not classified as ETGs, likely because of their distorted morphologies, possibly as the result of mergers. The fact that our RS cluster candidates are all at $z < 1.5$ seems to confirm that the redshift range of 3 to 1.5 is crucial for the formation of the RS and the evolution from proto-clusters to clusters of galaxies \citep{miley06, zirm08, galametz10}. 

Perhaps the most important result of this work is the discovery of the presence of a population of blue ETGs in the field of the most distant objects in the 3CR catalog. This is apparent in particular for 3C257 at a redshift z=2.47. The location of these early-type galaxies in the CMD is consistent with the predictions of our assumed simple passive evolution model for the redshift of the target. We also found evidence for a general trend with redshift for the color of the ETGs identified in all of our fields. The galaxies are typically bluer at higher redshifts, while they cluster around redder colors at z$<1.3$.
This implies that these ETGs include a significant component produced by a younger and thus bluer stellar population that is rapidly evolving towards the redder part of the diagram. Clearly such a finding must be further investigated using narrow band imaging centered on Ly$\alpha$ \citep[e.g][]{venemans04} and spectroscopy in order to both confirm these objects as cluster members and better study their stellar populations.

\acknowledgments{The authors thank Colin Norman for useful comments and insightful suggestions. J.P.K. and B.H. acknowledge support from $HST$-GO-13023.005-A. We thank the anonymous referee for her/his constructive comments that improved the paper. This research has made use of the NASA/IPAC Extragalactic Database (NED) which is operated by the Jet Propulsion Laboratory, California Institute of Technology, under contract with the National Aeronautics and Space Administration. This work is based in part on observations taken by the 3D-HST Treasury Program (GO12177 and 12328). This work is based on observations made with the NASA/ESA HST, obtained from the Data Archive at the Space Telescope Science Institute, which is operated by the Association of Universities for Research in Astronomy, Inc., under NASA contract NAS 5-26555.}

{\it Facilities:} \facility{HST (WFC3)}.

\clearpage
\end{document}